\documentclass[conference]{IEEEtran}
\IEEEoverridecommandlockouts
\usepackage{cite}
\usepackage{amsmath,amssymb,amsfonts}
\usepackage{algorithmic}
\usepackage{graphicx}
\usepackage{textcomp}
\usepackage{xcolor}

\usepackage{subcaption}

\def\BibTeX{{\rm B\kern-.05em{\sc i\kern-.025em b}\kern-.08em
    T\kern-.1667em\lower.7ex\hbox{E}\kern-.125emX}}
\begin{document}

\title{Geometry-Constrained EEG Channel Selection for Brain-Assisted Speech Enhancement}

\author{Keying Zuo, Qingtian Xu, Jie Zhang, Zhenhua Ling\\
NERC-SLIP, University of Science and Technology of China (USTC), Hefei, China\\
\{keyingzuo; qingtianxu\}@mail.ustc.edu.cn;
\{jzhang6; zhling\}@ustc.edu.cn
\thanks{K. Zuo and Q. Xu equally contributed to this work, which is financed by National Natural Science Foundation of China (62101523), Hefei Municipal Natural Science Foundation (2022012) and USTC Research Funds of the Double First-Class Initiative (YD2100002008). Correspondence: {\it Jie Zhang}.}}

\maketitle

\begin{abstract}
Brain-assisted speech enhancement (BASE) aims to extract the target speaker in complex multi-talker scenarios using electroencephalogram (EEG) signals as an assistive modality, as the auditory attention of the listener can be decoded from electroneurographic signals of the brain. This facilitates a potential integration of EEG electrodes with listening devices to improve the speech intelligibility of hearing-impaired listeners, which was shown by the recently-proposed BASEN model. As in general the multichannel EEG signals are highly correlated and some are even irrelevant to listening, blindly incorporating all EEG channels would lead to a high economic and computational cost. In this work, we therefore propose a geometry-constrained EEG channel selection approach for BASE. We design a new weighted multi-dilation temporal convolutional network (WD-TCN) as the backbone to replace the Conv-TasNet in BASEN. Given a raw channel set that is defined by the electrode geometry for feasible integration, we then propose a geometry-constrained convolutional regularization selection (GC-ConvRS) module for WD-TCN to find an informative EEG subset. Experimental results on a public dataset show the superiority of the proposed WD-TCN over BASEN. The GC-ConvRS can further refine the useful EEG subset subject to the geometry constraint, resulting in a better trade-off between performance and integration cost. 
\end{abstract}


\begin{figure*}[tbp]
\centerline{\includegraphics[width=0.98\textwidth]{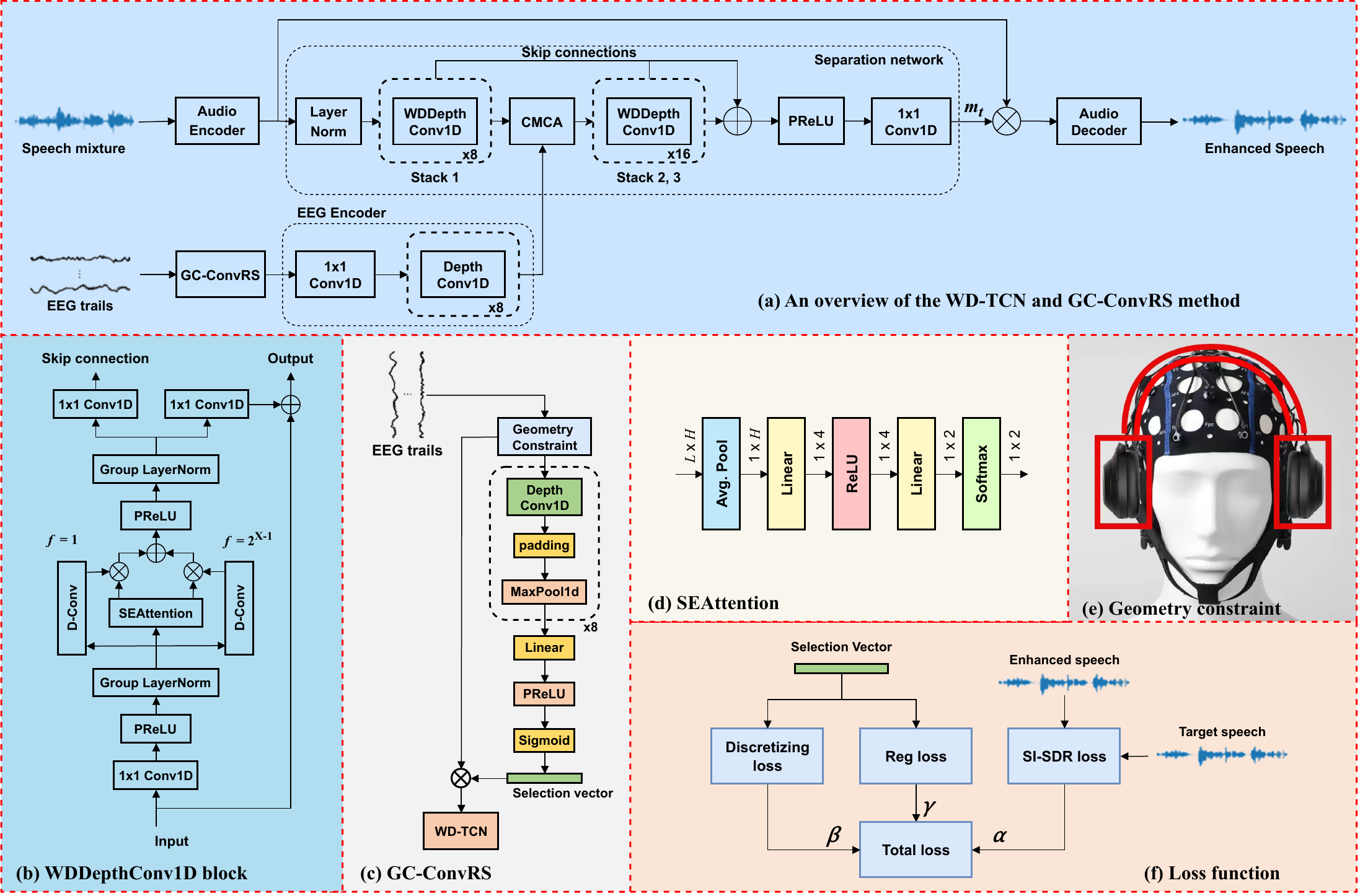}} % 这里将宽度设为列宽
\caption{The proposed GC-ConvRS sparsity-driven brain-assisted speech enhancement method: (a) The backbone WD-TCN model, (b) WDDepthConv1D block, (c) GC-ConvRS, (d) SEAttention and (e) Loss function.}
\label{fig1}
\vspace{-0.5cm}
\end{figure*}
\section{Introduction}
{Target speaker extraction (TSE) as a new branch of the classic speech enhancement (SE) aims to separate the target speech from noisy mixtures in multi-talker conditions, which has a wide range of applications in e.g., hearing aids (HAs), speech recognition, speaker diarization, remote conferencing, etc. Unlike blind source separation, the TSE usually requires additional clue(s) to specify the identity of the target speaker, which thus avoids the speaker permutation problem in speech separation. {\it The focus of this work is on the TSE for listening devices, e.g., HAs, in order to improve the speech quality and intelligibility for hearing-impaired listeners}.

Typical assistive clues for TSE may include 1) the direction of arrival (DOA) of the target speaker~\cite{wang2024study} that can be used to calculate the interchannel spatiotemporal features, 2) the pre-enrolled audio stream(s) uttered by the target speaker as  reference~\cite{ge2021multi} to help extract the speaker's voiceprint, and 3) the synchronized video containing  lip movements~\cite{pan2021muse} or face images~\cite{gao2021visualvoice,xu2023multi} that are associated with the target speaker. In practice, the DOA has to be estimated by other devices, which is heavily affected by dynamics, e.g., head movement. The pre-enrolled audio is often hardly available and background noises impact the extraction of speaker embeddings. The assistive video suffers from the illumination and occlusion conditions and is computationally more costly in case of integration with listening devices. {\it In this work, we thus seek electroencephalogram (EEG) signals to assist TSE, or more generally called \textbf{brain-assisted speech enhancement (BASE)}.}

%However, accessing these cues in practical use is still challenging, especially in the scenarios of listening devices. 

The human auditory system has the great ability to attend to the target speaker and it was shown that the target speaker can be decoded from the EEG signals, i.e., auditory attention decoding (AAD)~\cite{mesgarani2012selective,o2017neural,aroudi2020cognitive}, which facilitates the feasibility of using EEG signals for TSE. More importantly, the EEG cap with a certain amount of electrodes for data recording is physically natural for the integration with listening devices. Recently, some learning-based models have been proposed for the BASE task, e.g., non-end-to-end brain-informed speech separation (BISS)~\cite {ceolini2020brain} and end-to-end (E2E)  U-shaped brain-enhanced speech
denoiser (UBESD)~\cite{hosseini2021speaker,hosseini_end--end_2022}. The latter is more preferable with a better performance than the former. In~\cite{qiu2023tf}, a data augmentation-based TF-NSSE method was proposed, where the leakage of attention is futher analyzed. In~\cite{zhang2023basen}, an E2E BASEN model was proposed, which is based on the well-known Conv-TasNet~\cite{luo2019conv} backbone for speech extraction. More recently, NeuroHeed was proposed in~\cite{pan2023neuroheed} by incorporating the temporal association between the EEG signals and the attended speech and an improved version (NeuroHeed+) can be found in~\cite{pan2024neuroheed+}. The multi-scale fusion-based brain-assisted TSE network was proposed in~\cite{fan2024msfnet}, showing the usefulness of multi-scale features for BASE.} 

The aforementioned BASE approaches blindly include all EEG measurements for the fusion with the audio signal, which would cause a high hardware cost and setup time. In addition, practical listening devices do not allow the electrodes to be placed anywhere on the cap, where the acceptable integration should follow a headphone shape~\cite{zhang2023sparsity}.   {This raises an expectation of electrode placement via sparse channel selection~\cite{mirkovic2015decoding, strypsteen2021end, narayanan2019analysis,narayanan2020optimal}. Existing EEG channel selection mostly follows wrapper-based~\cite{liu2005toward,alotaiby2015review} and embedded approaches~\cite{strypsteen2021end}, where the latter is more preferable. A typical embedded method is the Gumbel channel selection (GCS)~\cite{strypsteen2021end}, which uses the Gumbel-softmax function~\cite{jang2016categorical} to approximate the selection status. The GCS suffers from the training difficulty in the combination with BASEN. This problem was solved in~\cite{xu2024end} using a residual Gumbel selection  and the duplicated selections were further reduced in~\cite{zhang2023sparsity}. However, the selection methods for BASE in~\cite{xu2024end,zhang2023sparsity} do not constrain the geometry  of the selected electrodes, which perform channel selection on the entire cap.} 

Based on the convolutional regularization selection (ConvRS) in~\cite{zhang2023sparsity}, in this work we therefore propose a more practical \textbf{G}eometry-\textbf{C}onstrained EEG channel selection method for BASE, called \textbf{GC-ConvRS}. Given the whole EEG channel set, we first consider a geometry constraint on the candidate allowable set that holds a headphone-like shape for the feasible integration with listening devices. This can be done by a manual selection.  We then propose a weighted multi-dilation temporal convolutional network (WD-TCN) to replace the Conv-TasNet backbone in BASEN~\cite{zhang2023basen}, which can thus dynamically focus more or less on the local information depending on the receptive field at each convolutional block~\cite{ravenscroft2022utterance}. The WD-TCN and GC-ConvRS are then combined to form the overall geometry-constrained BASE model. Experimental results on a public dataset show the superiority of the proposed GC-ConvRS method in two aspects: 1) the WD-TCD can achieve a better SE performance than BASEN~\cite{zhang2023basen}; 2) the selection of informative EEG channels subject to the geometry constraint decreases the performance, but not significantly, revealing the significance of EEG electrodes for speech perception as some are even irrelevant or contribute negatively. 

\section{Methodology}
The proposed GC-ConvRS method employs the WD-TCN as the BASE module to better leverage the temporal characteristics of speech signals for TSE and the ConvRS~\cite{zhang2023sparsity} to perform geometry-constrained EEG channel selection. Next, we will introduce each module in detail.

\subsection{WD-TCN}\label{WD-TCN}
As shown in Fig.~\ref{fig1}(a), the proposed WD-TCN is a {fully end-to-end time-domain BASE model, which is similar to the Conv-TasNet~\cite{luo2019conv} backbone in BASEN~\cite{zhang2023basen}}. A weighted multi-dilation depthwise-separable convolution is proposed to replace standard depthwise-separable convolutions in TCNs.  

Given the noisy speech signal $x$ and the matched $Q$-channel EEG signals $e_q,q=1,\cdots,Q$ recorded by the EEG cap worn by the listener, the audio encoder extracts the speech embedding sequence $w_x$ from $x$, and the EEG encoder estimates the EEG embedding $e_x$ from the recorded EEG trials. %That is,
%\begin{equation}
%    w_x = {\rm AudioEncoder}(x),  e_x = {\rm EEGencoder}(e_1,\cdots,e_K).
%\end{equation}
Both audio and EEG embeddings are input to the separation network to estimate the target speaker specific mask, given by
\begin{equation}
    m_t={\rm Separator}(w_x,e_x).
\end{equation}
%where $T$ denotes the number of existing sources. For target speaker extraction, $T$ = 2, i.e., the target speech and non-target components, while for the general speech separation task, $T$ might be greater than 2. 
Finally, the reconstructed speech signal is obtained by mapping the audio embedding using the source-specific masks, given by
\begin{equation}
    \hat{s}_t = {\rm Decoder}(w_x \odot m_t),
\end{equation}
where $\odot$ denotes the element-wise multiplication.

The audio encoder includes  a
couple of Depth-wise 1D convolutions (DepthConv1D) to extract the audio feature, and the decoder holds a mirror structure.
The EEG encoder consists of one 1D convolution to downsample the EEG signals and a couple of DepthConv1D, where each has residual connections to form multi-level EEG features. The Depth-wise convolution was shown to be effective in many tasks~\cite{kaiser2017depthwise, chollet2017xception}. The separator mainly consists of several weighted multi-dilation depthwise-separable convolutions (WDDepthConv1D) and the convolutional multi-layer cross attention (CMCA) module~\cite{zhang2023basen} for bi-modal feature fusion. The WDDepthConv1D block, see Fig.~\ref{fig1}(b), enables  the network to be more selective on the temporal context without drastically increasing the parameter amount. We design a squeeze-and-excite attention block~\cite{chen2020dynamic, hu2018squeeze}, see Fig.~\ref{fig1}(d), to compute the weights for each layer of conventional depthwise-separable convolution. %The squeeze-and-excite attention is shown in .

\subsection{GC-ConvRS}\label{GC-ConvRS}
On the basis of the WD-TCN backbone, we then consider a two-step geometry-constrained channel selection. 1) \textbf{Hard Selector:} We define an headphone-like area (see Fig.~1(e)) that the electrodes are allowed to be placed and determine a candidate set $\mathcal{S}$ from $\mathcal{Q}=\{1,\cdots,Q\}$. This is called hard pre-selector, which can be done manually and only depends on the user-defined geometry constraint. As such, the final selected channels can always satisfy a headphone-like integration of EEG electrodes with listening devices. 2) \textbf{Soft Selector:} We then apply the ConvRS method~\cite{zhang2023sparsity} to the candidate set $\mathcal{S}$ and refine the selection results. This step is thus called \textbf{soft refinement}. The structure of  ConvRS is shown in Fig.~\ref{fig1}(c), which {is composed of a series of DepthConv1D blocks and an linear layer to obtain the selection vector $\mathbf{s}$.} These two steps result in the proposed GC-ConvRS channel selection approach.

Specifically, the ConvRS considers two regularization loss functions to {constrain the discreteness and cardinality of the selection vector, which are shown in Fig.~\ref{fig1}(f)}. The discretization loss function can be written as~\cite{zhang2023sparsity}
\begin{equation}\label{eq:loss_Ld}
    \mathcal{L}_{d}=k_{1}\left(-\frac{\mathbf{d}^T \mathbf{d}}{QB} + b\right),
\end{equation}
where $B$ denotes the batch size, the bias $b$ equals 0.25 to set the minima
of $\mathcal{L}_{d}$ to be 0, and
%$k_{1}$ is a {\color{blue}scaling coefficient, which is set to 100}, and 
the divergence vector $\mathbf{d}=\mathbf{s}-0.5$. In order to decrease the number of the selected electrodes, we set a  regularization loss, given by
\begin{equation}
    \mathcal{L}_{\rm reg}=k_{2}||\mathbf{s}||_2^2.
\end{equation}
%where $k_{2}$ is a similar scaling balance parameter.%, which is set to 0.25.
In addition, the scale-invariant signal-to-distortion ratio (SI-SDR)~\cite{le2019sdr} is often-used to measure the speech quality:
\begin{equation}
    \mathcal{L}_{\rm SI\text{-}SDR}=10\log_{10}{\left \| x_\text{target} \right \|^2  } /{\left \| x_\text{res} \right \|^2  },
\end{equation}
where $x_\text{target}=\frac{\hat{s}_t^Ts}{\Vert s\Vert ^2}s$ and $x_\text{res}=x_\text{target}-\hat{s}_t$ with $s$ denoting the target speech. The negative SI-SDR is taken as the TSE-related loss function for model training.  %It was shown that SI-SDR is a satisfactory optimizing target for the time-domain SE~\cite{kolbaek2020loss}.}

As a result, the total loss function can be formulated as
\begin{equation}\label{eq:total-loss}
    \mathcal{L}=-\alpha \mathcal{L}_{\rm SI\text{-}SDR}+\beta \mathcal{L}_{d}+\gamma \mathcal{L}_{\rm reg},
\end{equation}
where $\alpha$, $\beta$ and $\gamma$ are weights of the three components. In this work, both $\alpha$ and $\beta$ are set to 0.5, but $\gamma$ is set depending on the expected amount of EEG channels, i.e., $\gamma$ controls the sparsity of the selected channel subset. Besides, we empirically set the scaling balance parameters as $k_1$ = 100 and $k_2$ = 0.25.

\section{Performance Evaluation}
\begin{figure*}[htbp]
  \centering
  \begin{subfigure}[b]{0.49\textwidth}
    \centering
    \includegraphics[width=\textwidth]{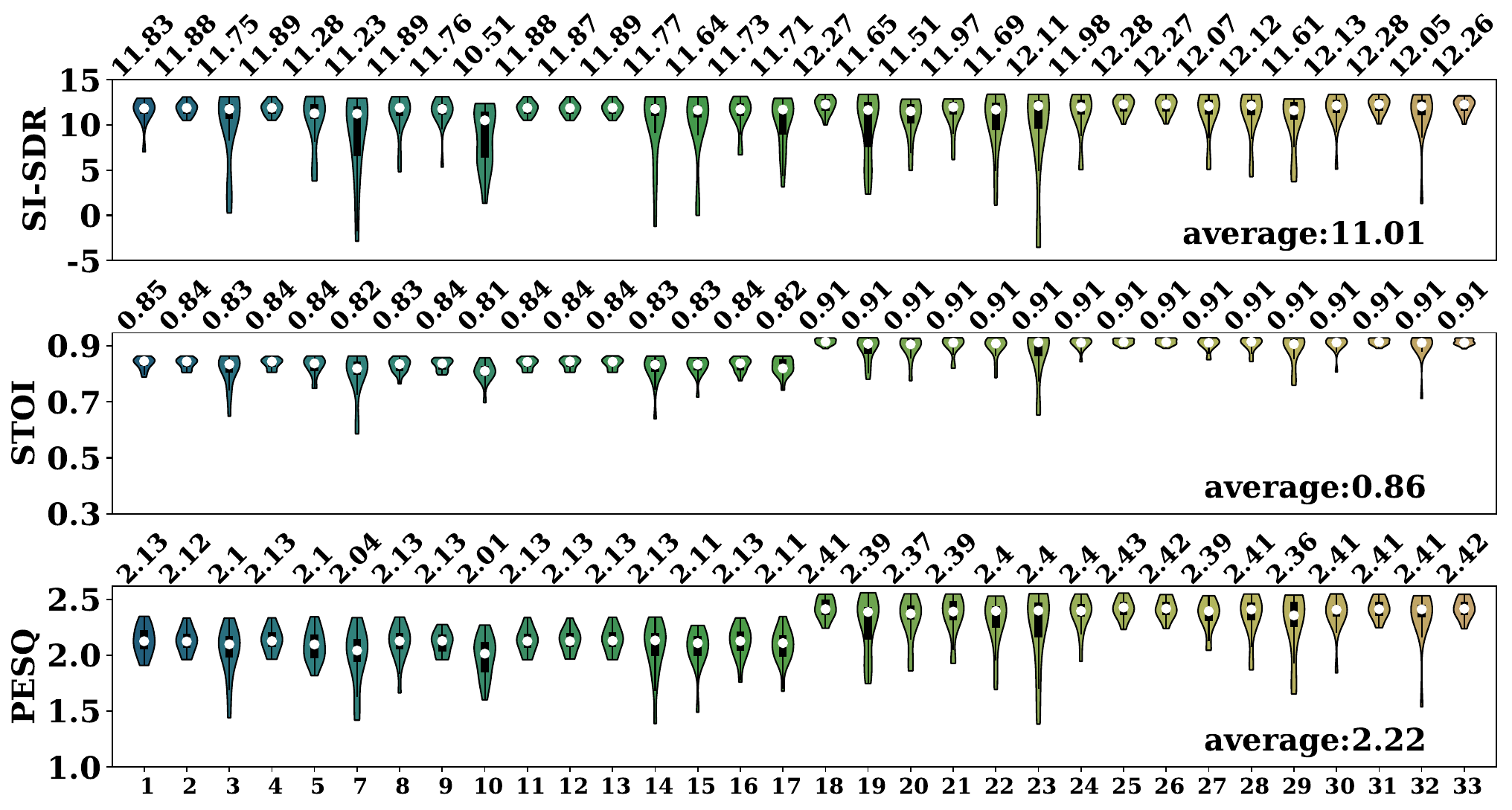}
    \caption{BASEN}
    \label{fig2:subfig1}
  \end{subfigure}
  \hfill
  \begin{subfigure}[b]{0.49\textwidth}
    \centering
    \includegraphics[width=\textwidth]{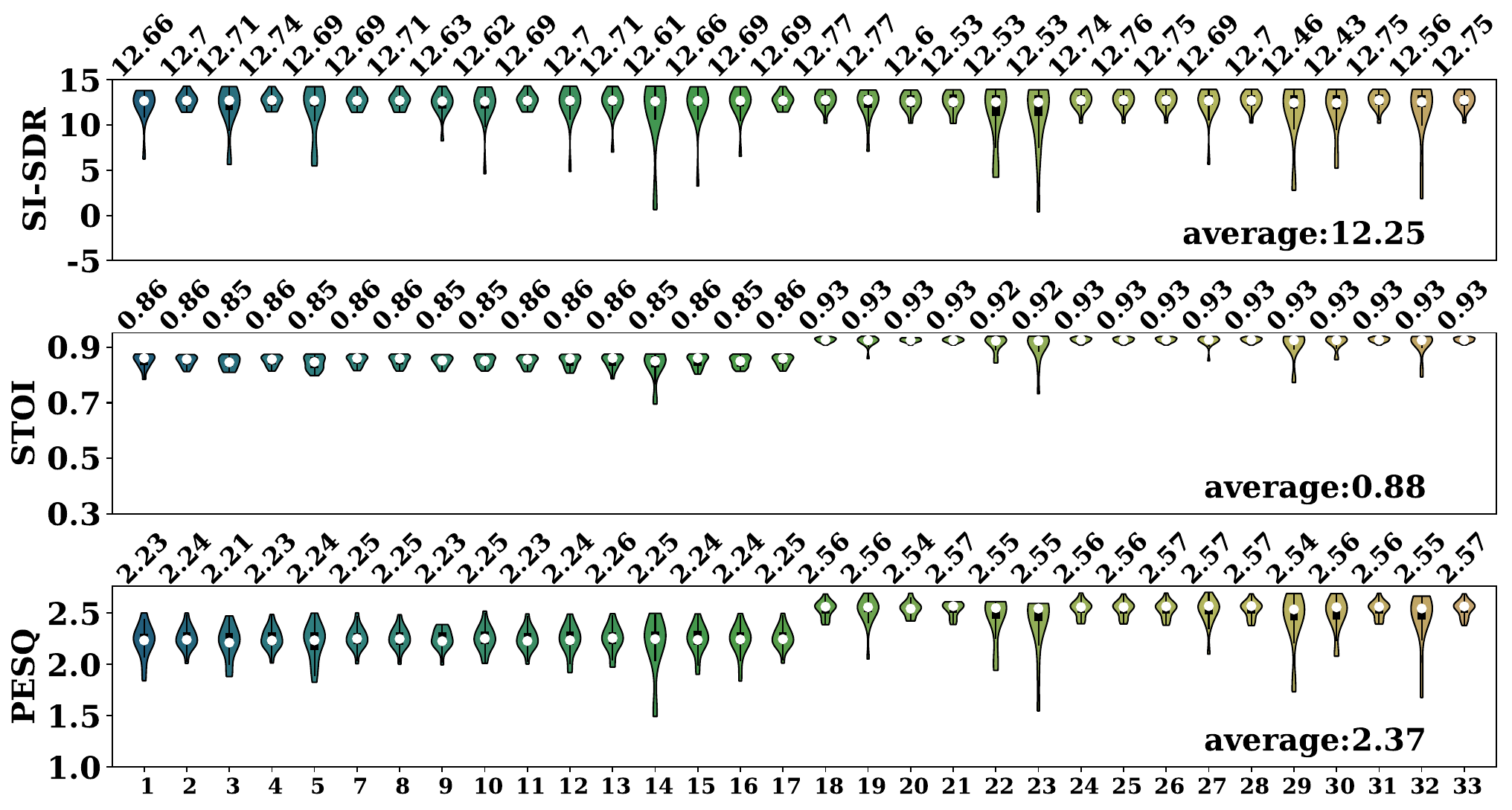}
    \caption{WD-TCN}
    \label{fig2:subfig2}
  \end{subfigure}
  \vspace{-0.05cm}
  \caption{Performance of  WD-TCN and BASEN~\cite{zhang2023basen} across subjects, where the median values are shown at the top of plots.}
  \label{fig2}
  \vspace{-0.3cm}
\end{figure*}

\begin{figure*}[htbp]
  \centering
  \begin{subfigure}[b]{0.49\textwidth}
    \centering
    \includegraphics[width=\textwidth]{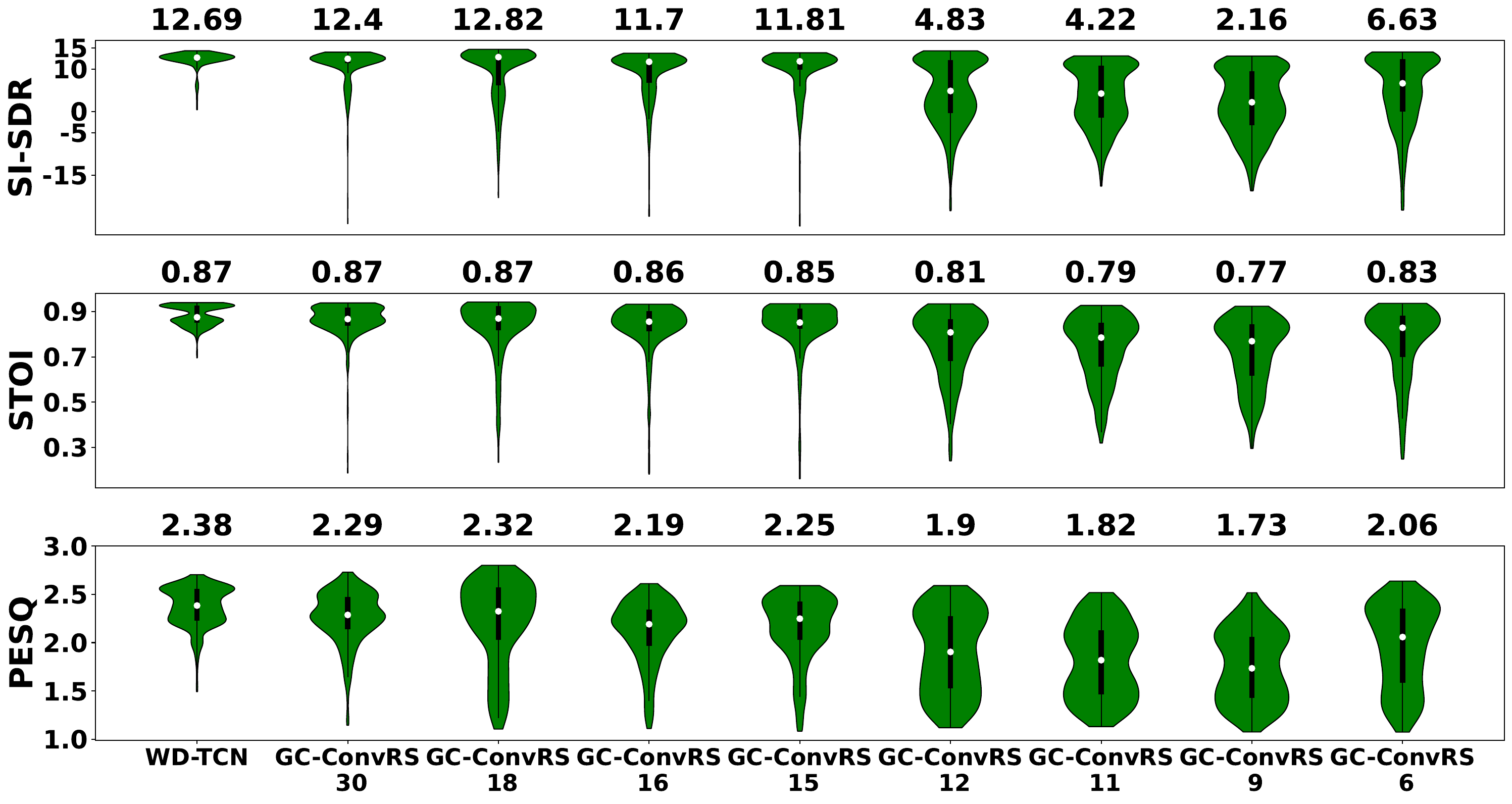}
    \caption{The performance of GC-ConvRS}
    \label{fig3:subfig1}
  \end{subfigure}
  \hfill
  \begin{subfigure}[b]{0.49\textwidth}
    \centering
    \includegraphics[width=0.85\textwidth]{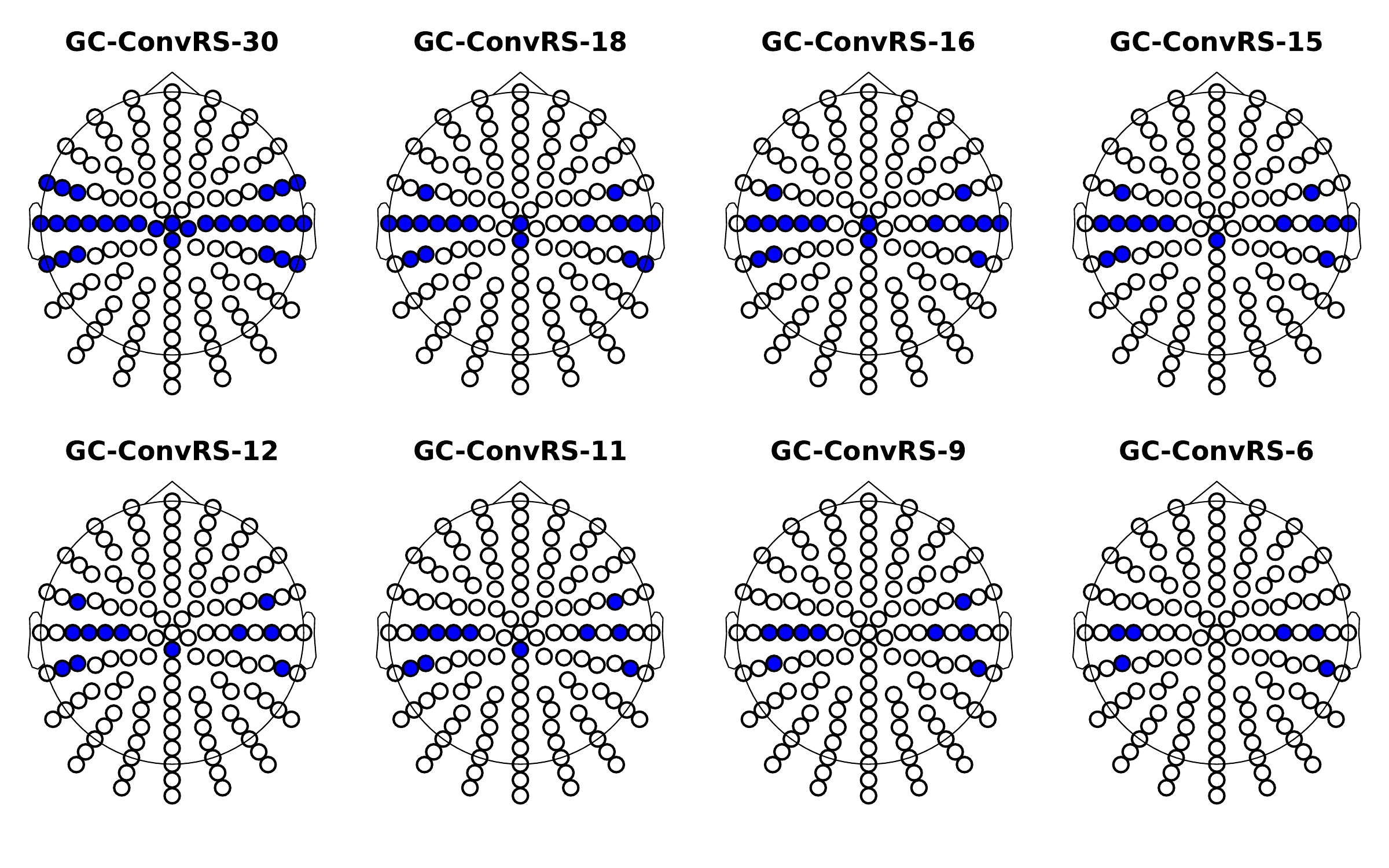}
    \caption{Visualization of the selected EEG channels}
    \label{fig3:subfig2}
  \end{subfigure}
    \vspace{-0.05cm}
  \caption{The performance of and visualization of GC-ConvRS, where the blue dots represent the selected electrodes.}
  \label{fig3}
  \vspace{-0.5cm}
\end{figure*}
\subsection{Experimental Setup}
\textbf{Dataset:} {In this work, we evaluate the BASE performance on the public dataset originated from \cite{broderick2018electrophysiological}, which includes 33 subjects. We exclude subject 6 due to the worse data quality. Each subject undertook 30 trials and each trial lasted 1 minute. In each trial, two stories were played to different ears. Half of the subjects were asked to attend to the left ear while the others to the right. That is, the attended story is regarded as the target speaker and the other as noise. EEG data were recorded using a 128-channel (plus two mastoids) EEG cap at a rate of 512 Hz and further downsampled to 128 Hz. More details on the dataset description can be found in \cite{broderick2018electrophysiological}}.

\textbf{Pre-processing:} {Our experimental setup is the same as \cite{hosseini2021speaker,hosseini_end--end_2022,zhang2023basen,zhang2023sparsity,xu2024end}. The audio signals were downsampled to 14.7 kHz and the two stimuli were normalized to have the same RMS level and then added to generate the noisy mixture at a signal-to-noise ratio (SNR) of 0 dB.  The dataset is split into three groups: randomly choosing 5 trials from all subjects for testing, 2 trials for validation, and the rest for model training. That is, we divide subsets with respect to trials, and no overlap of trials occurs among training, validation and test subsets. Note that one can also split groups on subjects. For training and validation, each trial was cut into 2-sec segments, while for testing each trial was cut into 20-sec segments.}

The EEG data were first pre-processed by a band-pass filter with the pass band ranging from 0.1 Hz to 45 Hz. Then the EEG signals were re-referenced to the average of the mastoid channels. We simply applied the independent component analysis (ICA) function in EEGLAB~\cite{delorme2004eeglab} to remove artifacts. {For each subject, we dropped the trial that contains too much noise}.
Similarly to~\cite{hosseini_end--end_2022}, we used the frequency-band coupling model~\cite{whittingstall2009frequency} to estimate the audio-related neural activity from EEG signals, which can be represented by the cortical multi-unit neural activity (MUA) from EEG signals~\cite{whittingstall2009frequency,moinnereau2020frequency}.

\textbf{Implementation:} {Three objective metrics are used to measure the BASE performance}, including SI-SDR~\cite{le2019sdr} in dB, perceptual evaluation of speech quality (PESQ)~\cite{rix2001perceptual} and short-time objective intelligibility (STOI)~\cite{taal2010short}. The higher, the better. The Adam optimizer~\cite{kingma2014adam} was used for training with a momentum of $\beta_{1}$ = 0.9 and a denominator momentum of $\beta_{2}$ = 0.999. We used the linear warm-up following the cosine annealing learning rate schedule with a maximum learning rate of 1e-4 and a warm-up ratio of 5\%. The model was trained for around 200 epochs with a batch size of $B$ = 8. 

\subsection{Experimental Results}

% \begin{figure*}[htbp]
%   \centering
%   \begin{subfigure}[b]{0.48\textwidth}
%     \centering
%     \includegraphics[width=\textwidth]{BASEN.pdf}
%     \caption{BASEN}
%     \label{fig:subfig1}
%   \end{subfigure}
%   \hfill
%   \begin{subfigure}[b]{0.48\textwidth}
%     \centering
%     \includegraphics[width=\textwidth]{WDTCN.pdf}
%     \caption{WD-TCN}
%     \label{fig:subfig2}
%   \end{subfigure}
%   \caption{the comparison of the WD-TCN method with the BASEN method, where the median values are shown at the top of each sub-figure. The white dot in each plot shows the median. The black bar in the center of the violins shows the interquartile range (IQR). The thin black lines stretched from the bar show {\it first quartile $-1.5\times IQR$} and {\it third quartile} $+1.5\times IQR$, respectively.}
%   \label{fig2}
% \end{figure*}

First, we compare the proposed WD-TCN backbone model and BASEN~\cite{zhang2023basen} using all $Q$ = 128 channels in Fig.~\ref{fig2}, which shows the SI-SDR, PESQ and STOI across 32 subjects. Clearly, the WD-TCN outperforms BASEN in all metrics and over all subjects, and the performance variance of WD-TCN is smaller.  Note that the BASEN model  approximately has a parameter amount of 0.654M, while the proposed WD-TCN has around 0.688M parameters. Indeed, the WD-TCN differs from BASEN by replacing the DepthConv1D blocks with WDDepthConv1D counterparts, which are comparable in size. But the adopted WDDepthConv1D makes use of squeeze-and-excite attention that can help to better capture temporal features of speech signals. This explains the superiority of the proposed WD-TCN backbone over BASEN.

%\subsection{Evaluation of GC-ConvRS EEG Channel Selection}
Second, we validate the efficacy of the proposed GC-ConvRS channel selection approach in comparison with the full-channel based WD-TCN in Fig.~3(a). To do this, we set a geometry constraint in a headphone shape to perform the initial hard selection, resulting in $|\mathcal{S}|$ = 30 EEG channels as shown by \textbf{GC-ConvRS-30} in Fig.~3(b). The subsequent soft refinement selections of GC-ConvRS are obtained by adapting the sparsity regularizer $\gamma $ picked from $\left\{0, 0.1, 0.2, 0.3, 0.4, 0.5, 0.6 \right\}$, leading to selected subsets of $\left\{18, 16, 15, 12, 11, 9, 6 \right\}$ channels, respectively. 
The BASE performance and selection examples are shown in Fig.~3. It is interesting that in case $|\mathcal{S}|$ = 18, the SI-SDR is even higher, the STOI and PESQ scores only change slightly compared to the full-channel inclusion. This tells that even when $Q$ = 30, there are still some irrelevant channels to speech understanding. In case 18 $>|\mathcal{S}|>$ 6, the performance generally decreases monotonically in terms of the number of selected channels, but the SI-SDR changes more rapidly. In the extreme case of $|\mathcal{S}|$ = 6, the performance is surprisingly  better than that of GC-ConvRS (9-12), which are clearly not good choices for the integration of electrodes. Overall, \textbf{GC-ConvRS-18} achieves a better trade-off between the BASE performance and the hardware cost (in terms of the electrode amount). Finally, from the visualization of selections, we observe that the channels close to the ears and the left and right temporal cortex are more likely to be chosen, since these brain areas are more related to speech perception.

\section{Conclusion}
In this paper, we proposed the WD-TCN backbone for the BASE task and the GC-ConvRS for geometry-constrained EEG channel selection. It was shown that the proposed WD-TCN is stronger for BASE and the GC-ConvRS does not decrease the average performance too much compared to the full-channel incorporation under the geometry constraint. The obtained selections all satisfy the geometry shape. This validates that the multichannel EEG signals are highly correlated and some channels are marginal or even useless to BASE. The exclusion of these channels by GC-ConvRS is thus practically meaningful for the hardware integration of EEG electrodes with listening devices. As the considered dataset plays the same mixture to the two ears, which does not fit the natural listening case with interaural differences, we will investigate the brain-assisted binaural SE in the future. %Besides, channel selection and visualization can help to find out which channels and the corresponding brain areas that are relevant to spatial speech perception. Notice that channel selection would also cause a larger performance variation across subjects, so we will consider subject-adaptive EEG channel selection in future work. Also, the integration of EEG devices with hearing aids remains unsolved yet, where we will consider using EEG signals to assist joint binaural speech separation and spatial cues preservation.

\vfill

\pagebreak
\clearpage

\bibliographystyle{IEEEtran}
\bibliography{ref}

\vspace{12pt}

\end{document}